\documentclass[5p,sort&compress]{elsarticle}
\usepackage[T1]{fontenc}
\usepackage[latin9]{inputenc}
\usepackage{bm}
\usepackage{amsmath}
\usepackage{graphics}
\usepackage{graphicx}
\usepackage{epsfig}
\usepackage{amssymb}

\makeatletter
\usepackage[english]{babel}
\makeatother

\begin{document}
\begin{frontmatter}

\title{Scattering by flexural phonons in suspended
graphene under back gate induced strain}

\author[ICMM]{Héctor Ochoa}
\author[ICMM,CFP]{Eduardo V. Castro}
\author[NS]{M. I. Katsnelson}
\author[ICMM]{F. Guinea}

\address[ICMM]{Instituto de Ciencia de Materiales de Madrid, CSIC, Cantoblanco, E-28049 Madrid, Spain}
\address[CFP]{Centro de Fìsica do Porto, Rua do Campo Alegre 687, P-4169-007 Porto, Portugal}
\address[NS]{Radboud University Nijmegen, Institute for Molecules and Materials, NL-6525 AJ Nijmegen, The Netherlands}

%\ead{email@address}

%\ead[url]{http://www.home.page}

\begin{abstract}
We have studied electron scattering by out-of-plane (flexural) phonon modes in doped suspended graphene and its effect on charge transport. In the free-standing case (absence of strain) the flexural branch shows a quadratic dispersion relation, which becomes linear at long wavelength when the sample is under tension due to the rotation symmetry breaking. In the non-strained case, scattering by flexural phonons is the main limitation to electron mobility. This picture changes drastically when strains above $\bar{u}=10^{-4} n(10^{12}\,\text{cm}^{-2})$ are considered. Here we study in particular the case of back gate induced strain, and apply our theoretical findings to recent experiments in suspended graphene.
\end{abstract}
\end{frontmatter}

\section{Introduction}

Graphene is a novel two dimensional material whose low-temperature conductivity is comparable to that of conventional metals \cite{Morozov_etal}, despite much lower carrier concentrations. Interactions with the underlying substrate seem to be the main limitation to electron mobility, and recent experiments on suspended samples show a clear enhancement of mobility (more than one order of magnitude) at low temperatures \cite{Du_etal,Bolotin_etal_1,Bolotin_etal_2}.

In suspended graphene carbon atoms can oscillate in the out-of-plane direction leading to a new class of low-energy phonons, the flexural branch \cite{Landau_book,Nelson_book}. In the free standing case, these modes show a quadratic dispersion relation, so there is a high number of these low-energy phonons and the graphene sheet can be easily deformed in the out-of-plane direction. For this reason it can be expected that flexural phonons are the intrinsic strongly T-dependent scattering mechanism which ultimately limits mobility at room temperature \cite{Katsnelson_Geim}. However, since the scattering process always involves two flexural phonons, a membrane characteristic feature, its effect could be reduced, specially at low temperatures \cite{Mariani_VonOppen}.

In the present manuscript we analyse theoretically the contribution of flexural modes to the resistivity in suspended graphene samples. Our results suggest, indeed, that flexural phonons are the main source of resistivity in this kind of samples. We also show how this intrinsic limitation is reduced by the effect of strain. A quantitative treatment of back gate induced strain where graphene is considered as an elastic membrane with clamped edges is given.

\section{The model}

In order to describe long-wavelength acoustic phonons graphene can be seen as a two dimensional membrane whose elastic properties are described by the free energy \cite{Landau_book, Nelson_book}
\begin{equation}
\mathcal{F}=\frac{1}{2}\kappa\int dxdy(\nabla^{2}h)^{2}+\frac{1}{2}\int dxdy(\lambda u_{ii}^{2}+2\mu u_{ij}^{2}).\label{eq:fe}\end{equation}where $\kappa$ is the bending rigidity, $\lambda$ and $\mu$ are
Lam\'e coefficients, $h$ is the displacement in the out of plane
direction, and $u_{ij} = 1/2
\left[ \partial_i u_j + \partial_j u_i + ( \partial_i h ) (
\partial_j h ) \right]$ is the strain tensor.
Typical parameters for
graphene \cite{Zakharchenko_etal} are $\kappa \approx 1\,$eV, and $\mu
\approx 3 \lambda \approx 9\,$eV$\,$\AA$^{-2}$. The mass density is
$\rho = 7.6 \times 10^{-7}\,$Kg/m$^2$. The longitudinal and transverse in-plane phonons show the usual linear dispersion relation with sound velocities $v_L = \sqrt{\frac{\lambda+2\mu}{\rho}}
\approx 2.1 \times 10^4$m/s and $v_T = \sqrt{\frac{\mu}{\rho}}
\approx 1.4 \times 10^4$m/s. Flexural phonons have the dispersion
\begin{equation}
\omega_{\bf q}^F=\alpha \left|{\bf q}\right|^2
\label{eq:flexural}
\end{equation}
with $\alpha = \sqrt{\frac{\kappa}{\rho}} \approx 4.6 \times
10^{-7}$m$^2$/s. The quadratic dispersion relation is strictly valid in the absence of strain. At finite strain the dispersion relation of flexural phonons becomes linear at long-wavelength due to rotation symmetry breaking. Let us assume a slowly varying strain field $u_{ij} ( \mathbf{r} )$. The
dispersion in Eq.~\eqref{eq:flexural} is changed to:
\begin{equation}
\omega_{\bf q}^F ( \mathbf{r} )=| \mathbf{q} |
\sqrt{\frac{\kappa}{\rho} | \mathbf{q} |^2 + \frac{\lambda}{\rho}
u_{ii} ( \mathbf r ) + \frac{2 \mu}{\rho} u_{ij} ( \mathbf{r} )
\frac{q_i q_j}{| \mathbf{q} |^2}}
\label{eq:flexuralstr}
\end{equation}
In order to keep an analytical treatment we assume uniaxial strain ($u_{xx} \equiv \bar{u}$, and the rest of strain components zero), and drop the anisotropy in Eq.~\eqref{eq:flexuralstr} by considering the effective dispersion relation
\begin{equation}
\omega_{\bf q}^F=q\sqrt{\alpha^2q^2+\bar{u}v_L^2}.
\label{eq:flexuralstrapp}
\end{equation}

Long-wavelength phonons couple to electrons in the effective Dirac-like Hamiltonian \cite{review} through a scalar potential (diagonal in sublattice indices) called the deformation potential, which is associated to the lattice volume change and hence it can be written in terms of the trace of the strain tensor \cite{Suzuura_Ando,Manes}
\begin{equation}
V ( {\bf r}  )=g_0 \left[ u_{xx} ( \mathbf{r} ) + u_{yy} (
\mathbf{r} ) \right]
\end{equation} where $g_0\approx 20 - 30\,$eV \cite{Suzuura_Ando}. Phonons couple also to electrons through a vector potential associated to changes in bond length between carbon atoms, and whose components are related with the strain tensor as \cite{Manes,Vozmediano_etal}
\begin{equation}
{\bf A} ( \mathbf{r} )=\frac{\beta}{a} \left\{ \frac{1}{2}
\left[ u_{xx} ( \mathbf{r} ) - u_{yy} ( \mathbf{r} ) \right] ,
- u_{xy} ( \mathbf{r} ) \right\}
\end{equation} where $a \approx 1.4\,$\AA is the
distance between nearest carbon atoms, $\beta =-\partial \log ( t ) /
\partial \log ( a ) \approx 2-3$ \cite{Heeger_etal}, and $t \approx 3\,$eV is
the hopping between electrons in nearest carbon $\pi$ orbitals.

Quantizing the displacements fields in terms of the usual bosonic $a_{\vec{\bf q}}^{i=L,T,F}$ operators for phonons of momentum $\mathbf{q}$ we arrive at the interaction Hamiltonian. The term which couples electrons and flexural phonons reads
\begin{align}
{\cal H}_{e-ph}^F &=
\sum_{\mathbf{k},\mathbf{k}'} \sum_{\mathbf{q},\mathbf{q}'}
\left( a_{\mathbf{q}}^F + {a_{- \mathbf{q}}^F}^\dag \right) \left( a_{\mathbf{q}'}^F +
{a_{- \mathbf{q}'}^F}^\dag \right) \delta_{\mathbf{k}',\mathbf{k} - \mathbf{q} - \mathbf{q}'} \nonumber \\
&\times \left[\sum_{c=a,b}
V^F_{1,\mathbf{q},\mathbf{q}'} c^\dag_{\mathbf{k}} c_{\mathbf{k}'} +
\left( V^F_{2,\mathbf{q},\mathbf{q}'} a^\dag_{\mathbf{k}} b_{\mathbf{k}'}
+ h.c. \right)\right],
\label{eq:coupling}
\end{align}
where operators $a_{\mathbf{k}}^\dag$ and $b_{\mathbf{k}}^\dag$
create electrons in Bloch waves with momentum $\vec{\bf k}$ in the
$A$ and $B$ sublattices respectively. The matrix elements are
\begin{align}
V_{1,\mathbf{q},\mathbf{q}'}^{F} &= -\frac{g_0}{2\varepsilon(\mathbf{q}+\mathbf{q}')}qq'\cos(\phi-\phi')\frac{\hbar}{2\mathcal{V}\rho\sqrt{\omega_{\mathbf{q}}^{F}\omega_{\mathbf{q}'}^{F}}},\nonumber \\
V_{2,\mathbf{q},\mathbf{q}'}^{F} &= -v_{F}\frac{\hbar\beta}{a}\frac{1}{4}qq'e^{i(\phi-\phi')}\frac{\hbar}{2\mathcal{V}\rho\sqrt{\omega_{\mathbf{q}}^{F}\omega_{\mathbf{q}'}^{F}}} \label{eq:potential}
\end{align}
where $\phi_{\bf q} = \arctan \left( q_y / q_x \right)$ and $\mathcal{V}$ is the volume of the system. The effect of screening has been taken into account in the matrix elements of deformation potential through a Thomas-Fermi -like dielectric function $\varepsilon\left(\mathbf q\right)=1+\frac{e^2\mathcal{D}\left(E_F\right)}{2\epsilon_0q}$, where $\mathcal{D}\left(E_F\right)$ is the density of states at Fermi energy. Note that $g = g_0 / \varepsilon(k_F) \approx 3\,$eV in
agreement with recent \emph{ab initio} results \cite{Choi_etal}.

\section{Resistivity in the absence of strain }
\label{sec:notstrain}

From the linearized Boltzmann equation we can calculate the resistivity as $\varrho=\frac{2}{e^{2}v_F^{2}\mathcal{D}(E_{F})}\frac{1}{\tau(k_{F})}$, where $v_F \approx 10^6\,$m/s is the Fermi velocity. Our aim is to compute the inverse of the scattering time of quasiparticles, given by $\tau_{\mathbf{k}}^{-1} = \sum_{\mathbf{k}'}
(1-\cos\theta_{\mathbf{k},\mathbf{k}'}) {\cal W}_{\mathbf{k},\mathbf{k}'}$, where ${\cal W}_{\mathbf{k},\mathbf{k}'}$ is the scattering probability per unit time, which can be calculated through the Fermi's golden rule.
For scattering processes mediated by two flexural phonons,
within the quasi-elastic approximation, we obtain
\begin{align}
{\cal W}_{\mathbf{k},\mathbf{k}'} &= \frac{4 \pi}{\hbar}
\sum_{i=1,2} \sum_{\mathbf{q},\mathbf{q}'} \left| V^F_{i,\mathbf{q},\mathbf{q}'} \right|^2 f^{(i)}_{\mathbf{k},\mathbf{k}'}
 \times \nonumber \\
& \times n_{\mathbf{q}} ( n_{\mathbf{q}' } +1 )
\delta_{\mathbf{k}',\mathbf{k} - \mathbf{q} - \mathbf{q}'}
\delta \left(E_{\mathbf{k}} - E_{\mathbf{k}'} \right)
\label{rate}
\end{align}
where $f^{(1)}_{\bf k,\bf k'} = 1 + \cos\theta_{\mathbf{k},\mathbf{k}'}$
and $f^{(2)}_{\mathbf{k},\mathbf{k}'} = 1$,
$n_{\mathbf q}$ is the Bose distribution, and
$E_{\mathbf{k}} = v_F \hbar k$ is the quasi-particle
dispersion for the Dirac-like Hamiltonian \cite{review}.
Eq.~\eqref{rate} is valid in the high $T$ limit to be specified
in the following.

In order to obtain analytical expressions for the scattering rates it is useful to introduce the Bloch-Gr\"uneisen temperature $T_{BG}$. If we take into account that the relevant phonons which contribute to the resistivity are those of momenta $q\gtrsim 2 k_F$ then we have $k_B T_{BG}=\hbar \omega_{2k_F}$. For in-plane longitudinal (transverse) phonons $T_{BG}=57 \sqrt n \,\,\text{K}$ ($T_{BG}=38 \sqrt n \,\,\text{K}$), where $n$ is expressed in $10^{12}$ cm$^{-2}$. For flexural phonons in the absence of strain $T_{BG}=0.1n \,\,\text{K}$. From the last expression it is obvious that for carrier densities of interest the experimentally relevant regime is $T\gg T_{BG}$, so let us concentrate on this limit.

In the case of scattering by in-plane phonons at $T\gg T_{BG}$ the scattering
rate is given by \cite{us}
\begin{equation}
\frac{1}{\tau_I} \approx \left[ \frac{g^2}{2v_L^2} + \frac{\hbar^2
v_F^2 \beta^2}{4 a^2}\left(\frac{1}{v_L^2}+\frac{1}{v_T^2}\right) \right]
\frac{E_F}{2 \rho \hbar^3 v_F^2} k_B T, \label{eq:tau_inplane}
\end{equation}
where now $g \approx 3\,$eV is the screened deformation potential constant. At $T\ll T_{BG}$ the scattering rate behaves as $\tau^{-1}\sim T^4$, where only the gauge potential contribution is taken into account since the deformation potential is negligible in this regime due to screening effects ($\tau^{-1}\sim T^{6}$ \cite{Hwang_DasSarma}).

In the case of flexural phonons in the non-strained case (in practice
$\bar{u}\ll 10^{-4}n$ with $n$ in $10^{12}\,\text{cm}^{-2}$),
the scattering rate at $T\gg T_{BG}$ reads \cite{us}
\begin{align}
\frac{1}{\tau_{F}} &\approx \left( \frac{g^2}{2} + \frac{\hbar^2
v_F^2 \beta^2}{4 a^2} \right) \frac{( k_B T )^2}{64
\pi \hbar \kappa^2 E_F} \ln \left( \frac{k_B T}{\hbar \omega_c}
\right) + \nonumber \\ &+ \left( \frac{g^2}{4} + \frac{\hbar^2
v_F^2 \beta^2}{4 a^2} \right) \frac{k_B T E_F}{32 \pi
v_F^2 \kappa \sqrt{\rho \kappa}} \ln \left( \frac{k_B T}{\hbar
\omega_c} \right) \label{eq:tau_flexural}
\end{align}
where we have taken into account two contributions, one coming from the absorption or emission of two thermal phonons, and other involving one non-thermal phonon. The first one dominates over the second at $T\gg T_{BG}$. It is necessary to introduce an infrared cutoff frequency $\omega_c$, where for small but finite
strain $\bar{u}\ll 10^{-4}n(10^{12}\,\text{cm}^{-2})$ is just
the frequency below which the flexural phonon dispersion becomes linear.

\begin{figure}
\label{comp}
\begin{center}
\includegraphics[width=0.7\columnwidth]{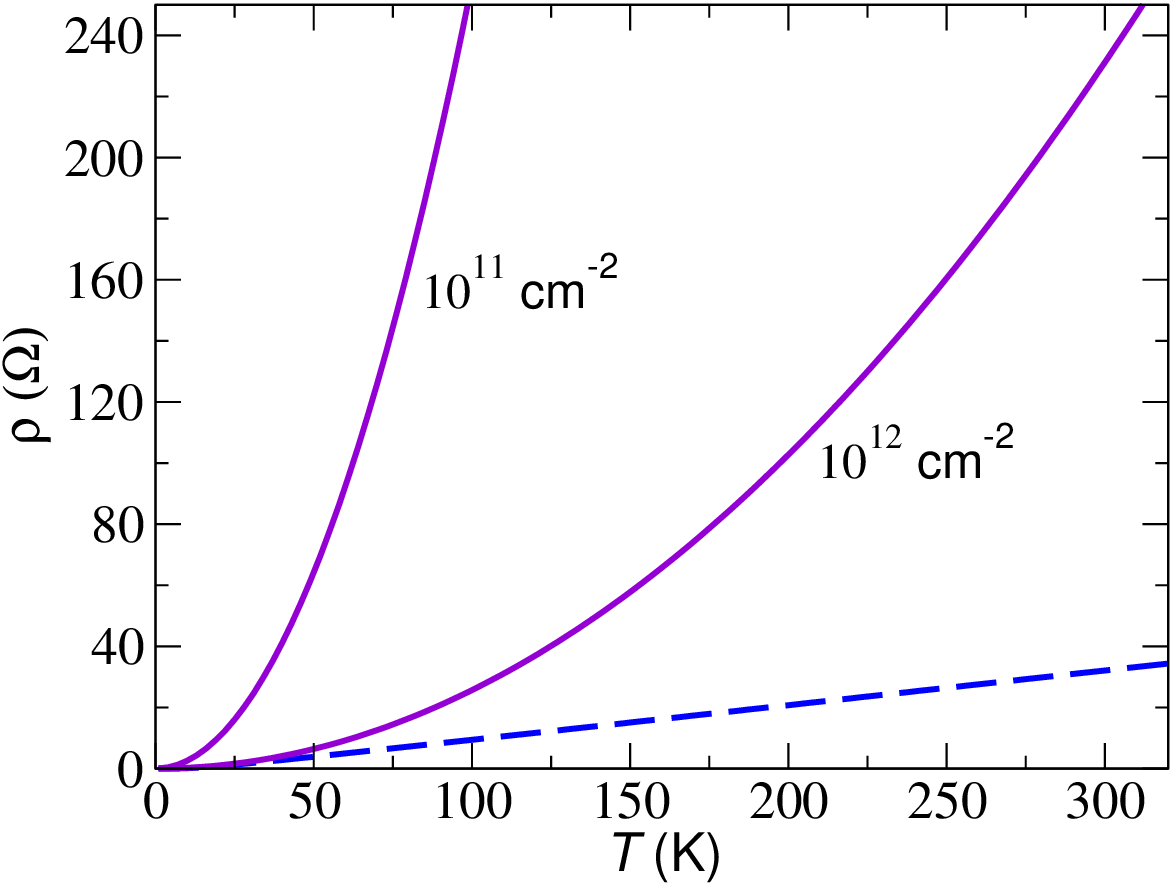}
\caption{Contribution to the resistivity
from flexural phonons in the absence of strains
for
two different electronic concentrations (full lines)
and from in plane phonons (dashed line).}
\end{center}
\end{figure}

From Eq.~\eqref{eq:tau_inplane} we deduce a resistivity which behaves
as $\varrho\sim T$, with no dependence on $n$, whereas from
Eq.~\eqref{eq:tau_flexural} we have
(neglecting the logarithmic correction) $\varrho\sim T^2/n$, as it was deduced for classical ripples in \cite{Katsnelson_Geim}. As it can be seen in
Fig.~\ref{comp}, the resistivity due to scattering by flexural phonons
dominates over the in-plane contribution. However, this picture changes
considerably if one considers strain above $10^{-4}n(10^{12}\,\text{cm}^{-2})$,
as is discussed in the next section.

\section{Resistivity at finite strains}
\label{sec:strain}

\subsection{Scattering rate}
\label{subsec:sr}

The Bloch-Gr\"uneisen temperature for flexural phonons at
finite strains $\bar{u} \gtrsim 10^{-4}n(10^{12}\,\text{cm}^{-2})$
is $T_{BG} = 28 \sqrt{\bar u n}\,\, \text{K}$. In the relevant high-temperature
regime, $T\gg T_{BG}$ the scattering rate can be written as \cite{us}
\begin{align}
\frac{1}{\tau_{F}^{str}} &\approx \left( \frac{g^2}{4} +
\frac{\hbar^2 v_F^2 \beta^2}{4 a^2} \right)
\frac{E_F
( k_B T )^4}{16\pi \rho^2 \hbar^5 v_F^2 v_L^6 \bar{u}^3}  \times \nonumber\\
&\times
 \left[\mathcal{R}_2\left(\frac{\alpha k_BT}{\hbar v_L^2 \bar{u}}\right) +
\mathcal{R}_1\left(\frac{\alpha k_BT}{\hbar v_L^2 \bar{u}} \right) \right]
  \label{eq:tau_flexural_2}
\end{align}
where $\mathcal{R}_{n}(\gamma)=\int_{0}^{\infty}dx\,\frac{x^{3}}{(\gamma^{2}x^{2}+1)[\exp(\sqrt{\gamma^{2}x^{4}+x^{2}})-1]^{n}}$. The two terms in Eq.~\eqref{eq:tau_flexural_2} come from the same processes as in Eq.~\eqref{eq:tau_flexural} described above. It is possible to obtain asymptotic analytical expressions for
Eq.~\eqref{eq:tau_flexural_2}. For instance, in the limit $T\ll T^* = \frac{\hbar v_L^2\bar{u}}{\alpha k_B}\approx 7\times 10^3 \bar{u}\,\,\text{K}$ the scattering rate behaves as $\tau^{-1}\sim \frac{T^4}{\bar{u}^3}$, whereas in the opposite limit it behaves as $\tau^{-1}\sim \frac{T^2}{\bar{u}}$. The temperature $T^*$
characterises the energy scale at which the flexural phonon dispersion under
strain Eq.~\eqref{eq:flexuralstrapp} cross over from linear to quadratic.

It is
pertinent to compute  the crossover temperature $T^{**}$ above which scattering
 by flexural phonons dominates when strain is induced. This can be
inferred by comparing Eq.~\eqref{eq:tau_inplane} with
Eq.~\eqref{eq:tau_flexural_2} and imposing
$\tau_{I}/\tau_F \approx 1$. The numerical solution give for the
corresponding crossover $T^{**} \approx 10^6\bar{u}\,\, {\rm K}$.
Since $T^{**} \gg T^*$ we can use the respective asymptotic expression for
Eq.~\eqref{eq:tau_flexural_2}, $\tau^{-1}\sim \frac{T^2}{\bar{u}}$ to obtain
$T^{**} \approx 32 \pi \kappa \bar u /k_B \approx 10^6\bar{u}\,\, {\rm K}$.
A remarkable conclusion may then be drawn: scattering due to flexural
phonons can be completely suppressed by applying strain as low as
$\bar u \gtrsim 0.1\% $.

\subsection{Back gate induced strain}

\begin{figure}
\begin{center}
\includegraphics[width=0.7\columnwidth]{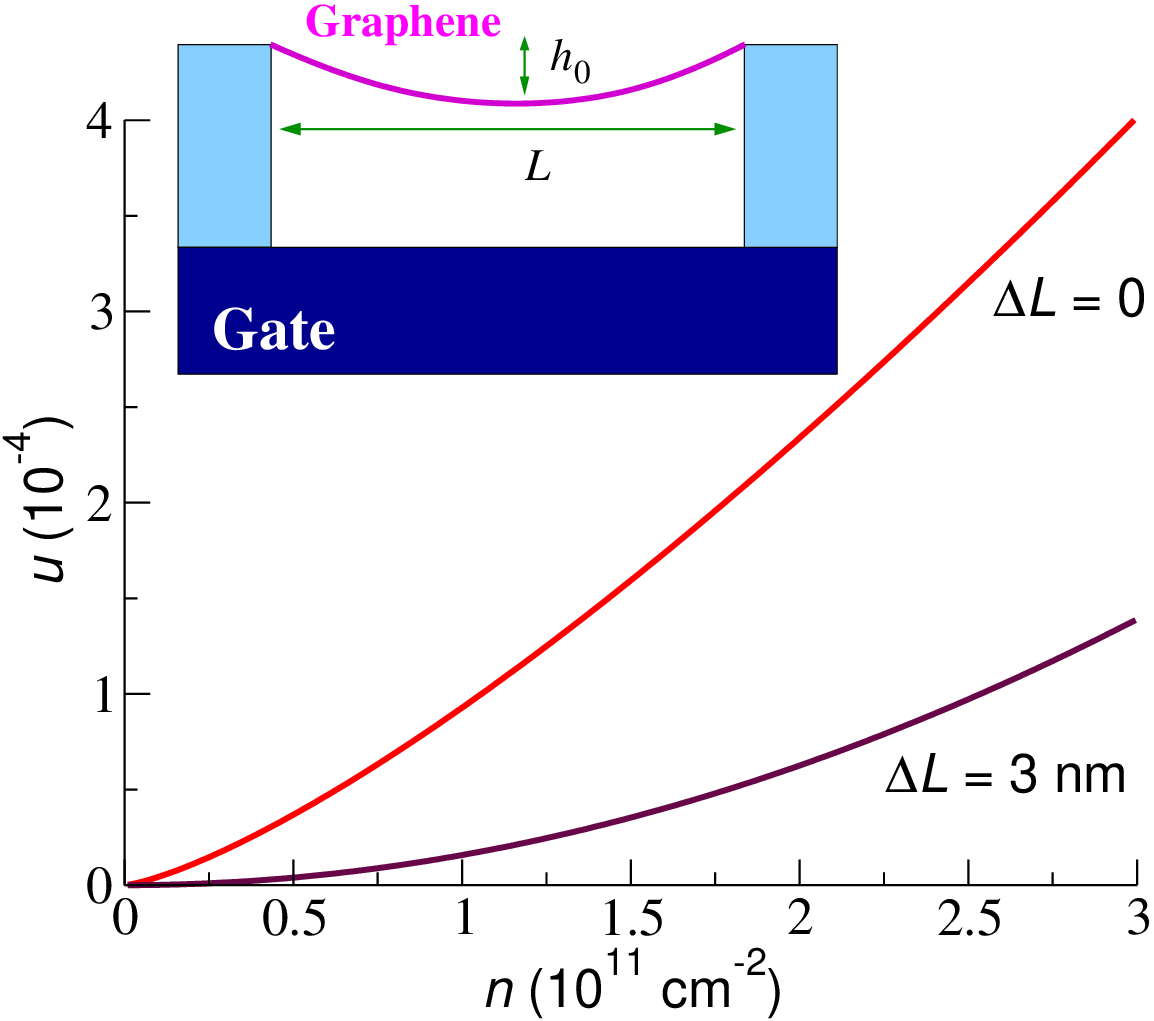}
\caption[fig]{Strain induced by the back gate in a suspended graphene
membrane of length $L= 1\,\,\mu m$ as a function
of the respective carrier density for two different $\Delta L$ (slack).
Inset: sketch of a suspended graphene membrane with clamped edges.}
\end{center}
\label{fig:strain}
\end{figure}

In order to compute the strain induced by the back gate we consider
the simplest case of a suspended membrane with clamped edges. A side
view of the system is given in the inset of Fig.~\ref{fig:strain}.

The static height profile is obtained by minimising the free energy,
Eq.~\eqref{eq:fe} in the presence of the load
$P=e^{2}n^{2}/(2\varepsilon_{0})$ due to the back gate induced electric field.
The built up strain is related with the applied load as
\cite{Landau_book,Fogler},
\begin{equation}
\bar{u}=\frac{PL^{2}}{8h_{0}(\lambda+2\mu)}
\approx5\times10^{-5}\frac{[n(10^{12}\mbox{cm}^{-2})
L(\mu\mbox{m})]^{2}}{h_{0}(\mu\mbox{m})},
\label{eq:uP}
\end{equation}
where $L$ is the length of the trench over which graphene is clamped
and $h_{0}$ is the maximum deflection (see the inset of Fig.~\ref{fig:strain}).
We assume the length of the suspended graphene region in the undeformed case
to be $L + \Delta L$, where the $\Delta L$ can be either positive or negative.
Under the approximation
of nearly parabolic deformation (which can be shown to be the relevant
case here \cite{Fogler}) the maximum deflection $h_{0}$ is given by
the positive root of the cubic equation
\begin{equation}
\left(h_{0}^{2}-\frac{3}{8}L\Delta L\right)h_{0} =
\frac{3PL^{4}}{64(\lambda+2\mu)},
\label{eq:root}
\end{equation}
with trench/suspended-region length mismatch $\Delta L$ such that
$\Delta L\ll L$.
If $\Delta L=0$ then Eq.~\eqref{eq:root} can be easily solved and we obtain for strain
\begin{equation}
\bar{u}=\frac{1}{2\sqrt[3]{3}}\left(\frac{PL}{\lambda+2\mu}\right)^{2/3}
\approx2\times10^{-3}(n^{2}L)^{2/3},
\label{eq:strainNoSlack}
\end{equation}
with $n$ in $10^{12}\,\mbox{cm}^{-2}$ and $L$ in $\mu\mbox{m}$.

In Fig.~\ref{fig:strain} the back gate induced strain is plotted as
a function of the respective carrier density. For a typical density
$n\sim 10^{11}\, \text{cm}^{-2}$  and $\Delta L =0$ we see that a back
gate induces strain $\bar u \sim 10^{-4}$. This would imply a crossover
from in-plane dominated resistivity $\varrho \sim T$
to $\varrho \sim T^2 /\bar u$ due to flexural phonons
at $T^{**} \sim 100\,\,\text{K}$, well within experimental reach.
In the next section we will argue that the experimental data
in Ref.~\cite{Bolotin_etal_2} can be understood within this framework.
Note, however, that gated samples can also fall in the category of
non-strained system if $\Delta L > 0$. This is clearly seen
in Fig.~\ref{fig:strain} for $\Delta L$ as small as
$\Delta L / L \approx 0.3\%$.

\subsection{Resistivity estimates: comparison with experiment}

\begin{figure}
\begin{center}
\includegraphics[width=0.9\columnwidth]{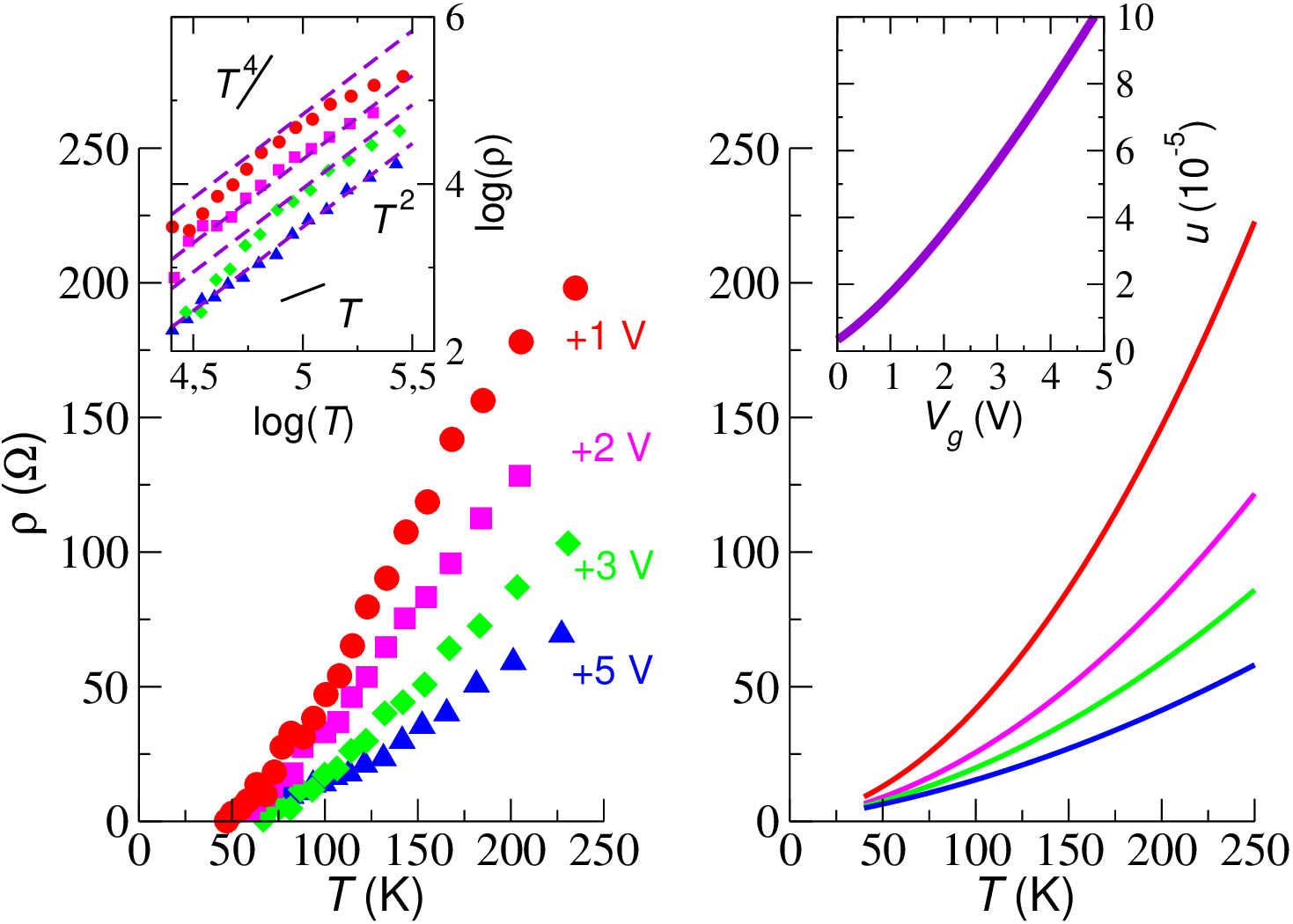}
\caption[fig]{Left: Temperature dependent resistivity from
Ref.~\cite{Bolotin_etal_2} at different gate voltages; the inset
shows the same in log log scale. Right: Result of Eq.~\ref{eq:rhofit};
the inset shows the back gate induced strain as given by Eq.~\eqref{eq:uP}.}
\end{center}
\label{fig:expfit}
\end{figure}

Bolotin et al. \cite{Bolotin_etal_2} have recently measured
the temperature dependent resistivity in doped suspended graphene.
The experimental results are shown in the left panel of Fig.~\ref{fig:expfit}
in linear scale, and the inset shows the same in log log scale.
In Ref.~\cite{Bolotin_etal_2} the resistivity was interpreted
as linearly dependent on temperature for $T\gtrsim 50\,\,\text{K}$.
In the left inset of Fig.~\ref{fig:expfit}, however, it becomes apparent
that the behaviour is closer to the $T^2$ dependence in the high temperature
regime (notice the slopes of $T^4$ and $T$ indicated in full lines and that
of $T^2$ indicated as dashed lines). Within the present framework the obvious
candidates to explain the quadratic temperature dependence are flexural
phonons. Since the measured resistivity is too small to be due to scattering
by non-strained flexural phonons we are left with the case of flexural
phonons under strain, where the strain can be naturally assigned to the back
gate.

In the right panel of Fig.~\ref{fig:expfit} we show the theoretical
$T-$dependence of the resistivity taking into account scattering
by in-plane phonons and flexural phonons with finite strain,
\begin{equation}
\varrho = \frac{2}{e^{2}v_F^{2}\mathcal{D}(E_{F})}
\left(\frac{1}{\tau_I} + \frac{1}{\tau_F^{str}}\right),
\label{eq:rhofit}
\end{equation}
where $1/\tau_I$ is given by Eq.~\eqref{eq:tau_inplane} and $1/\tau_F^{str}$
by Eq.~\eqref{eq:tau_flexural_2}. We calculated the back gate induced strain
via Eq.~\eqref{eq:strainNoSlack}, and related the density and gate
voltage as in a parallel plate capacitor model, $n\simeq C_g(V_g-V_{NP})/e$
\cite{Bolotin_etal_1, Bolotin_etal_2} ($C_g = 60\,\, \text{aF}/\mu \text{m}^2$
and $V_{NP} \approx -0.4\,\,\text{V}$). The obtained strain is shown in the
right inset of Fig.~\ref{fig:expfit} versus applied gate voltage. It is seen
that the system is well in the region where
Eq.~\ref{eq:tau_flexural_2} is valid. The agreement between left and right
panels in Fig.~\ref{fig:expfit} for realistic parameter values \cite{params}
 is an indication that we are indeed observing
the consequences of scattering by flexural phonons at finite, though very small
strains. Full quantitative agreement is not aimed, however, since our two
side clamped membrane is a very crude approximation to the real device
\cite{Bolotin_etal_1, Bolotin_etal_2}.

\section{Conclusions}

Our theoretical results suggest that scattering by flexural phonons constitute the main limitation to electron mobility in doped suspended graphene. This picture changes drastically when the sample is strained. In that case, strains with not too large values, as those induced by the back gate, can suppress significantly this source of scattering. This result opens the door to the possibility of modify locally the resistivity of a suspended graphene by strain modulation.


\begin{thebibliography}{00}
\bibitem{us} Derivation details will be given elsewhere.
\bibitem{params} We used $g = 3\, \text{eV}$, $\beta = 3$,
$\Delta L = 0$, and $L = 0.3\,\mu\text{m}$. The latter parameter is interpreted
as an effective length mimicking the difference between our two side clamped
membrane and the real four point clamped device.
\end{thebibliography}
\end{document}